# Evolution of Anisotropic Displacement Parameters and Superconductivity with Chemical Pressure in BiS$_2$-Based REO$_{0.5}$F$_{0.5}$BiS$_2$ (RE = La, Ce, Pr, and Nd)

Yoshikazu Mizuguchi[1], Kazuhisa Hoshi[1], Yosuke Goto[1], Akira Miura[2], Kiyoharu Tadanaga[2], Chikako Moriyoshi[3], Yoshihiro Kuroiwa[3]

1. Graduate School of Science and Engineering, Tokyo Metropolitan University, 1-1, Minami-osawa, Hachioji 192-0397, Japan.
2. Faculty of Engineering, Hokkaido University, Kita-13, Nishi-8, Kita-ku, Sapporo, Hokkaido 060-8628, Japan.
3. Department of Physical Science, Hiroshima University, 1-3-1 Kagamiyama, Higashihiroshima, Hiroshima 739-8526, Japan.

* Corresponding author: Yoshikazu Mizuguchi (mizugu@tmu.ac.jp)

Abstract

In order to understand the mechanisms behind the emergence of superconductivity by the chemical pressure effect in REO$_{0.5}$F$_{0.5}$BiS$_2$ (RE = La, Ce, Pr, and Nd), where bulk superconductivity is induced by the substitutions with a smaller-radius RE, we performed synchrotron powder X-ray diffraction, and analyzed the crystal structure and anisotropic displacement parameters. With the decrease of the RE$^{3+}$ ionic radius, the in-plane disorder of the S1 sites significantly decreased, very similar to the trend observed in the Se-substituted systems: LaO$_{0.5}$F$_{0.5}$BiS$_{2-x}$Se$_x$ and Eu$_{0.5}$La$_{0.5}$FBiS$_{2-x}$Se$_x$. Therefore, the emergence of bulk superconductivity upon the suppression of the in-plane disorder at the chalcogen sites is a universal scenario for the BiCh$_2$-based superconductors. In addition, we indicated that the amplitude of vibration along the $c$-axis of the in-plane chalcogen sites may be related to the $T_c$ in the BiCh$_2$-based superconductors.

BiS$_2$-based layered compounds have been extensively studied owing to the discovery of superconductivity in electron-doped phases and their relatively high transition temperatures ($T_c$), as high as 11 K [1–3]. A typical parent phase of the BiS$_2$-based superconductor is REOBiS$_2$ (RE: rare earth element), which has a layered structure composed of alternate stacks of REO insulating (blocking) layers and BiS$_2$ electrically conducting layers; it has a band gap of ~1 eV [2–5]. Electron carriers can be introduced in the BiS$_2$ layers by partial substitutions of the elements of the blocking REO layers [6–11]. For example, in LaO$_{1-x}$F$_x$BiS$_2$, the partial substitutions of O$^{2-}$ sites with F$^-$ generate electron carriers; the carrier concentration can be manipulated by varying the amount of substituted F [2]. Bulk superconductivity with a large shielding fraction in the magnetization measurements was not observed in electron-doped LaO$_{1-x}$F$_x$BiS$_2$ samples prepared using the solid-state reaction method at ambient pressure; however, filamentary superconducting states with a small shielding volume fraction were observed at $T_c$ ~2.5 K [2]. In addition, the observed temperature dependence of the electrical resistivity was not metallic-like; it showed a weakly localized behavior in LaO$_{1-x}$F$_x$BiS$_2$, although band calculations suggested that the electron-doped LaO$_{1-x}$F$_x$BiS$_2$ should be metal [4, 5, 12]. These results suggested that the doped electrons were localized by the effect of structural disorder.

In order to induce bulk superconductivity in LaO$_{1-x}$F$_x$BiS$_2$, high pressure effects can be employed. The application of an external pressure induces bulk superconductivity with a $T_c$ of ~10 K [13–15]. In addition, samples annealed under a high pressure (~2 GPa) also exhibit bulk superconductivity with a $T_c$ of ~10 K [2, 16–18]. The emergence of bulk superconductivity and increase of $T_c$ in the high-pressure phase can be attributed to the structural phase transition from the tetragonal low-$T_c$ phase to the monoclinic high-$T_c$ phase [13].

Another approach to induce bulk superconductivity in the LaO$_{1-x}$F$_x$BiS$_2$ system is to introduce a chemical pressure by an isovalent substitution, such as Ch (Ch: S, Se) and/or RE site substitutions. In LaO$_{0.5}$F$_{0.5}$BiS$_{2-x}$Se$_x$, the substitutions of Se for the S sites induces bulk superconductivity with a $T_c$ of ~3.8 K [19, 20]. Another isovalent substitution is the RE site substitution. With the decrease of the RE$^{3+}$ (mean) ionic radius in REO$_{0.5}$F$_{0.5}$BiS$_2$, the BiS$_2$ layer becomes compressed, in particular along the *ab*-plane direction, and bulk superconductivity is induced [21]. As the isovalent substitution does not significantly affect the carrier concentration, the structural optimization induces superconductivity in the Ch- and RE-substituted systems. In order to analyze the essential factor for the emergence of the superconductivity, we considered the commonality of the chemical pressure effects between the Ch- and RE substitutions, by introducing the concept of an in-plane chemical pressure [22].

The emergence of the bulk superconductivity in the LaO$_{0.5}$F$_{0.5}$BiS$_{2-x}$Se$_x$ system was explained by the decrease in the in-plane disorder of the chalcogen sites with the increase of the Se content, which was detected through a crystal structure analysis of the synchrotron X-ray diffraction (SXRD) pattern [23] and extended X-ray absorption fine structure (EXAFS) [24]. A similar relationship between the in-plane disorder and emergence of superconductivity was observed in Eu$_{0.5}$La$_{0.5}$FBiS$_{2-x}$Se$_x$ [25–27].

For both systems, the Se substitutions induced a metallic electrical conductivity. Therefore, the in-plane disorder of the in-plane chalcogen sites is the factor that prevented the emergence of superconductivity and metallic conductivity in $LaO_{1-x}F_xBiS_2$. Therefore, we expect that a similar suppression of the in-plane disorder at the in-plane S sites (S1 site in Fig. 1(b)) can be attributed with the emergence of superconductivity in the RE-substituted systems. Neutron diffraction and pair distribution function analysis revealed the presence of the in-plane disorder and its relationship with the superconductivity, where RE is La or Nd [28, 29]. In addition, EXAFS measurements reveled a suppression of the Bi-S1-bond-disorder with the decrease of the $RE^{3+}$ radius [30]. Therefore, in this study, we investigated the atomic displacement parameters of $REO_{0.5}F_{0.5}BiS_2$ with RE = La, Ce, Pr, and Nd using SXRD and Rietveld refinement. Anisotropic analyses of the displacement parameters revealed that the in-plane displacement of S1 was very large for RE = La and gradually decreased with the decrease of the $RE^{3+}$ ionic radius. This behavior is equal to that of the Se substitution. Therefore, the suppression of the in-plane disorder at the in-plane S1 (or Ch1) sites and emergence of superconductivity by the chemical pressure effects are universal for $BiCh_2$-based superconductors. As unconventional superconductivity states are likely to appear in both Ch-substituted $La(O,F)Bi(S,Se)_2$ and RE-substituted $REO_{0.5}F_{0.5}BiS_2$ systems [31–33], the universality revealed in this study could be employed to further understand the mechanisms of the superconductivity in the $BiCh_2$-based superconductors.

Polycrystalline samples of $REO_{0.5}F_{0.5}BiS_2$ with RE = La, Ce, Pr, and Nd were prepared using the solid-state-reaction method [2, 6–9]. A mixture of the starting materials was pressed into a pellet and annealed at 700 °C for 20 h in an evacuated quartz tube. SXRD was performed at the beamline BL02B2, SPring-8 at a wavelength of 0.49559 Å (proposal No.: 2017B1211). The SXRD experiments were performed with a sample rotator system at room temperature; the diffraction data were collected using a high-resolution one-dimensional semiconductor detector (multiple MYTHEN system [34]) with a step size of $2\theta = 0.006°$. The crystal structure parameters were refined using the Rietveld method with the RIETAN-FP program [35]. Images with schematic models of the crystal structures were produced using VESTA [36].

Figure 1(a) shows the superconductivity phase diagram of $REO_{0.5}F_{0.5}BiS_2$ as a function of the $RE^{3+}$ ionic radius. The filamentary superconducting states observed for La are suppressed for Ce. A polycrystalline sample of $CeO_{0.5}F_{0.5}BiS_2$ with 50% F substitution does not show superconductivity; however, $CeO_{0.3}F_{0.7}BiS_2$ with 70% F substitution shows bulk superconductivity upon the effect of high pressure [37]. As reported in Ref. 21, bulk superconductivity of $REO_{0.5}F_{0.5}BiS_2$ appears at an $RE^{3+}$-radius of ~113 pm, which corresponds to RE = Pr or $Ce_{0.5}Nd_{0.5}$. $T_c$ increases with the decrease of the $RE^{3+}$ radius. The temperature dependences of the magnetization for the samples used in this study are shown in Figure S1 of the Supplemental Material [38]. $T_c$ values estimated from the magnetization measurements are represented with the larger circles in Fig. 1(a), which are in agreement with the results of a previous study [21].

Figure 2 shows the SXRD patterns and Rietveld refinement results for RE = (a) La, (b) Ce, (c) Pr, and (d) Nd. The Rietveld refinements were performed using a tetragonal *P4/nmm* model. For RE = La, Ce, and Pr, a small portion (smaller than 4%) of $REF_3$ impurity phases was detected, and a two-phase analysis was performed. For RE = Nd, unidentified impurity peaks were observed; hence, we analyzed the data for RE = Nd by a single-phase analysis. For all data, excellent fitting was also obtained at higher angles, required for a precise refinement of the displacement parameters.

Figures 3(a) and 3(b) show the dependences of the lattice constants *a* and *c* as a function of the $RE^{3+}$ ionic radius. The lattice constant *a* linearly decreases with the decrease of the $RE^{3+}$ radius. However, the lattice constant *c* does not exhibit a clear correlation with the $RE^{3+}$ radius. Figure 3(c) shows the dependences of the Bi-S1 and Bi-S2 bond distances as a function of the $RE^{3+}$ ionic radius (See Fig. 1(b) for the labeling of the S sites). It can be noticed that the in-plane Bi-S1 distance decreases with the decrease of the $RE^{3+}$ radius, owing to the linear decrease in the lattice constant *a*. In contrast, Bi-S2 and interplane Bi-S1 distances do not decrease upon the introduction of chemical pressure, which may be related with the nonlinear change in the lattice constant *c*. However, as reported in Ref. 22, the essential factor for the emergence of superconductivity in the $REO_{0.5}F_{0.5}BiCh_2$ systems is the in-plane chemical pressure. In the considered $REO_{0.5}F_{0.5}BiS_2$ system, the amplitude of the in-plane chemical pressure is directly determined by the in-plane Bi-S1 distance. Therefore, we can analyze the relationship between the chemical pressure, superconductivity, and in-plane disorder using the considered samples.

We performed Rietveld refinements with anisotropic displacement parameters $U_{11}$ and $U_{33}$ for the in-plane Bi and S1 sites. Figure 4(c) illustrates $U_{11}$ and $U_{33}$ using a schematic model with RE = Pr. For the other sites, isotropic displacement parameters were assumed for the refinements. The obtained $U_{11}$ and $U_{33}$ for the Bi and S1 sites are plotted in Figs. 4(a) and 4(b), respectively. As expected, $U_{11}$ for the S1 sites is very large for RE = La, and decreases with the decrease of the $RE^{3+}$ radius, a very similar trend with that observed for $U_{11}$ for the in-plane chalcogen sites in $LaO_{0.5}F_{0.5}BiS_{2-x}Se_x$ [23]. Therefore, the large suppression of the in-plane disorder at the in-plane chalcogen sites can be universally attributed with the emergence of bulk superconductivity in the $REO_{0.5}F_{0.5}BiCh_2$ systems. It is worth noting that $U_{11}$ for the Bi sites slightly increases with the decrease of the $RE^{3+}$ radius. Recently, Athauda et al. proposed that the increase in the in-plane displacement of Bi may be attributed with the superconductivity in $NdO_{1-x}F_xBiS_2$, while the in-plane displacement of S1 would induce carrier localization [29]. Our analysis also suggests that the increase of the in-plane displacement of the Bi sites may be related with the enhanced $T_c$ for RE = Pr and Nd. Similar trend is observed for $U_{33}$ in Fig. 4(b). Although $U_{33}$ for the Bi sites does not considerably change with the decrease of the $RE^{3+}$ radius, $U_{33}$ for the S1 sites dramatically changes with the replacement of the RE. It is worth noting that the evolution of $U_{33}$ for the S1 sites seems to be linked with the enhancement of $T_c$. A similar evolution of the displacement ellipsoids is observed in Se-substituted $LaO_{0.5}F_{0.5}BiSSe$, as shown in Fig. 5. Both RE and Ch substitutions suppress $U_{11}$ for the Ch1 sites; however, they enhance $U_{33}$ for the Ch1 sites. This trend may suggest that the amplitude of the one-dimensional large vibration of S1 is correlated with the $T_c$ in this system. If this is valid, then the $T_c$ increases when the phonon frequency decreases with the increase of the amplitude of the S1 vibration along the $c$-axis. This scenario is not consistent with the conventional phonon-mediated mechanisms. It is consistent with the isotope effect; no isotope effect was observed for the $LaO_{0.6}F_{0.4}Bi(S,Se)_2$ samples with $^{76}Se$ and $^{80}Se$ isotopes [31]. If such large vibrations of Ch1 are related with the enhancement of $T_c$, charge fluctuation should occur in the superconducting Bi-Ch1 plane. Then, the pairing phenomena mediated by charge fluctuations, proposed by theoretical and experimental studies [28, 29, 32, 39], may occur in the $BiCh_2$-based system.

In conclusion, we investigated the crystal structure and anisotropic displacement parameters of $REO_{0.5}F_{0.5}BiS_2$ with RE = La, Ce, Pr, and Nd, where bulk superconductivity is induced by substitutions with a smaller-radius RE, such as Pr or Nd. With the decrease of the $RE^{3+}$ ionic radius, the lattice constant $a$ and in-plane Bi-S1 distance monotonically decreased, which generated the in-plane chemical pressure effect. With the decrease of the $RE^{3+}$ radius, the in-plane disorder of the S1 sites significantly decreased. This trend is very similar to those observed for Se-substituted $LaO_{0.5}F_{0.5}BiS_{2-x}Se_x$ and $Eu_{0.5}La_{0.5}FBiS_{2-x}Se_x$. Therefore, the emergence of bulk superconductivity upon the suppression of the in-plane disorder at the Ch1 sites seems to be a universal scenario for the $BiCh_2$-based superconductors. The analyses of the displacement parameters along the $c$-axis indicated that the amplitude of the one-dimensional vibration of S1 (or Ch1) along the $c$-axis was correlated with $T_c$ in the $BiCh_2$-based superconductors.

Acknowledgements
We would like to thank O. Miura for his experimental support. This study was partially supported by the Grants-in-Aid for Scientific Research (Nos. 15H05886, 16H04493, 16K17944, 17K19058, and 17H04950) and JST-CREST (No. JPMJCR16Q6), Japan.

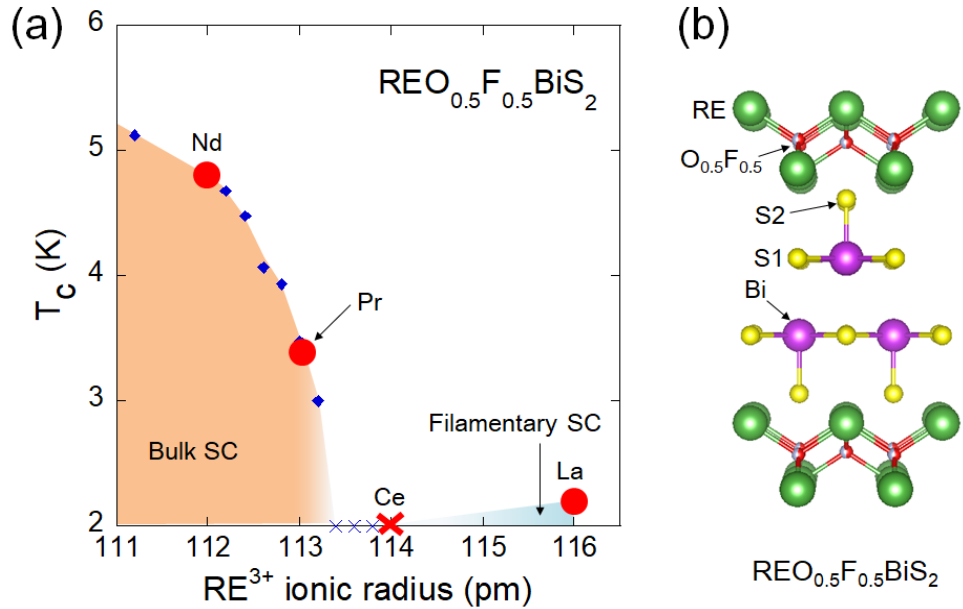

Fig. 1. (Color online) (a) Superconductivity phase diagram of $REO_{0.5}F_{0.5}BiS_2$ as a function of the $RE^{3+}$ ionic radius with a coordination number of 6. Data points for $Ce_{1-x}Nd_xO_{0.5}F_{0.5}BiS_2$ and $Nd_{1-x}Sm_xO_{0.5}F_{0.5}BiS_2$ have been published in Ref. 26. Bulk and filamentary SC denote the bulk and filamentary superconducting states with weak diamagnetic signals, respectively. (b) Schematic model of the crystal structure of $REO_{0.5}F_{0.5}BiS_2$.

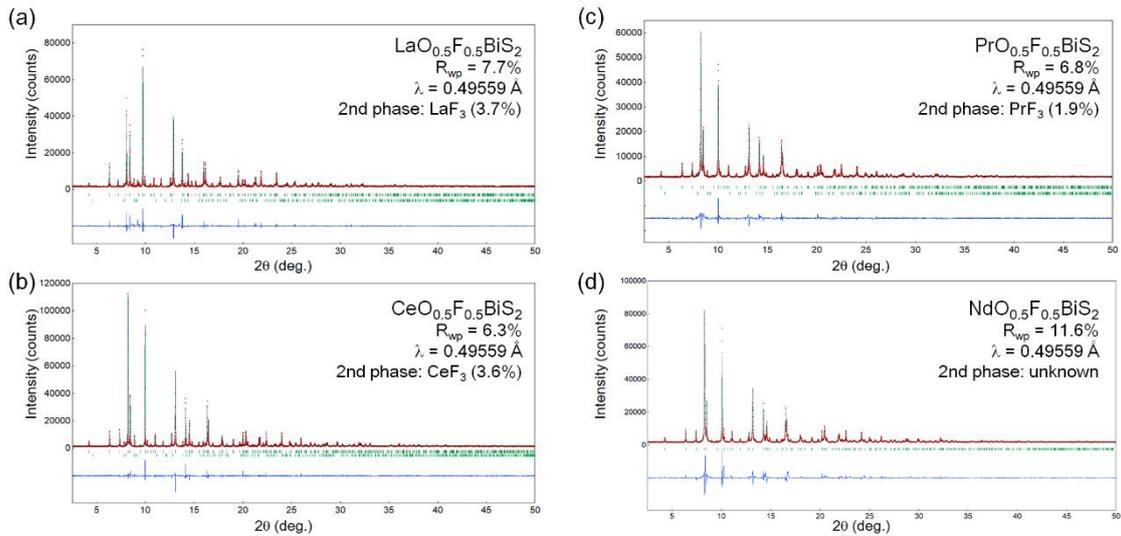

Fig. 2. (Color online) SXRD patterns for (a) $LaO_{0.5}F_{0.5}BiS_2$, (b) $CeO_{0.5}F_{0.5}BiS_2$, (c) $PrO_{0.5}F_{0.5}BiS_2$, and (d) $NdO_{0.5}F_{0.5}BiS_2$.

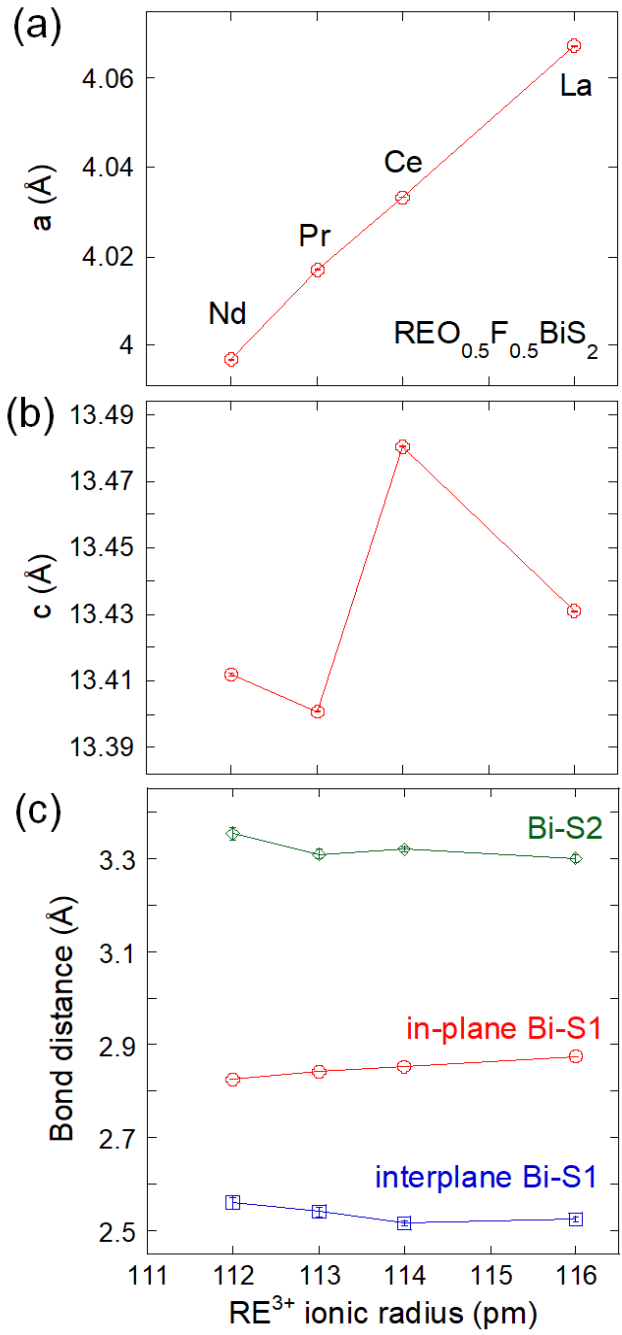

Fig. 3. (Color online) Dependences of the lattice constants (a) *a* and (b) *c*, and (c) Bi-S bond distances, as a function of the RE$^{3+}$ ionic radius.

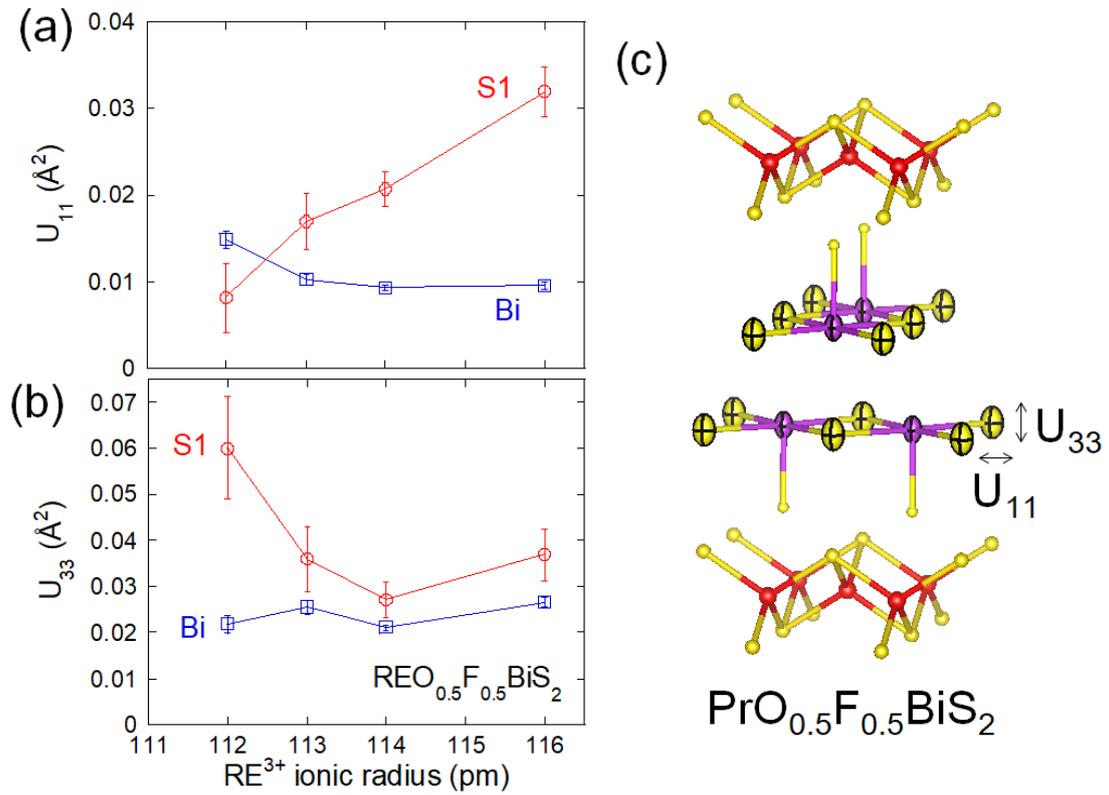

Fig. 4. (Color online) Dependences of the anisotropic displacement parameters (a) $U_{11}$ and (b) $U_{33}$ for the Bi and S1 sites as a function of the $RE^{3+}$ ionic radius. (c) Schematic model of the crystal structure of $PrO_{0.5}F_{0.5}BiS_2$; 90%-probability displacement ellipsoids are outlined.

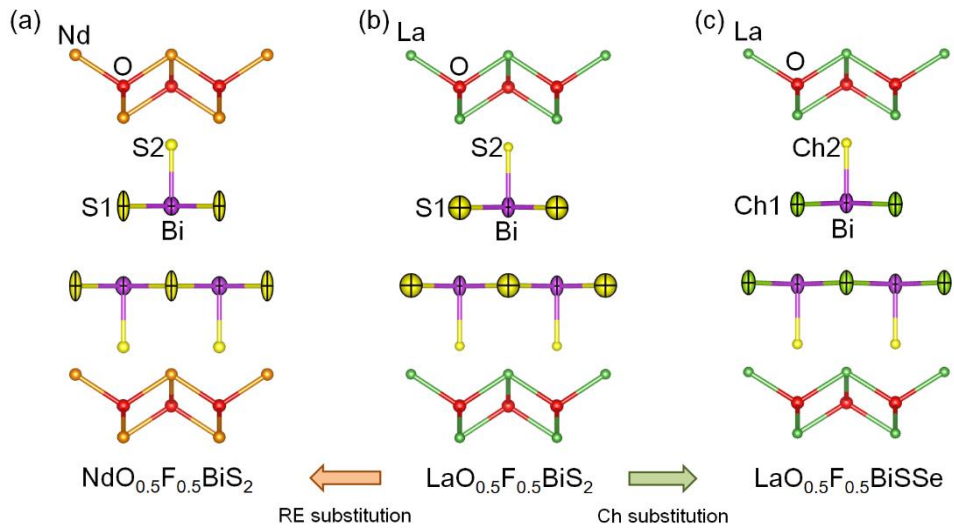

Fig. 5. (Color online) Evolution of the displacement ellipsoids (90% probability) by the effect of chemical pressure (RE or Ch substitutions) in the $REO_{0.5}F_{0.5}BiCh_2$ system. For $LaO_{0.5}F_{0.5}BiSSe$, Se selectively occupies the Ch1 sites.

# Supplemental Materials

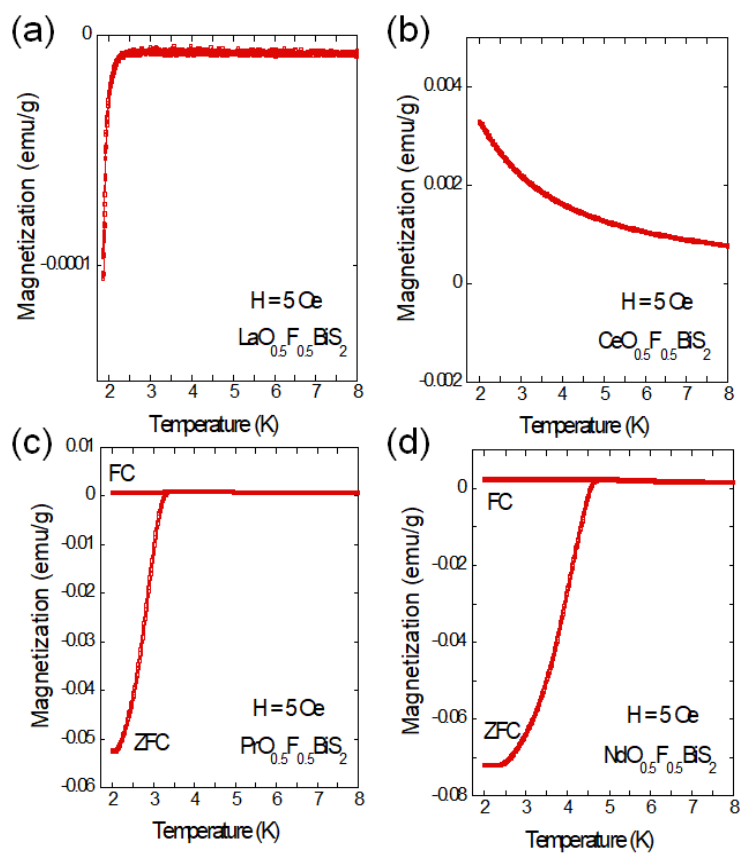

Fig. S1. Temperature dependence of the magnetization for (a) $LaO_{0.5}F_{0.5}BiS_2$, (b) $CeO_{0.5}F_{0.5}BiS_2$, (c) $PrO_{0.5}F_{0.5}BiS_2$, and (d) $NdO_{0.5}F_{0.5}BiS_2$.

Table SI. Crystal structure parameters for REO$_{0.5}$F$_{0.5}$BiS$_2$ obtained by the Rietveld refinements. The atomic coordinate for the oxygen (fluorine) site is (0, 0, 0), and the $U_{iso}$ is fixed as 0.013 Å$^2$. The atomic coordinate for other sites is (0, 0.5, $z$).

| RE | La | Ce | Pr | Nd |
|---|---|---|---|---|
| Space group | P4/*nmm* | P4/*nmm* | P4/*nmm* | P4/*nmm* |
| $a$ (Å) | 4.06732(4) | 4.03329(2) | 4.01708(6) | 3.99671(8) |
| $c$ (Å) | 13.4310(2) | 13.48059(10) | 13.4008(3) | 13.4118(4) |
| $V$ (Å$^3$) | 222.189(4) | 219.294(3) | 216.247(6) | 214.236(8) |
| $R_{wp}$ (%) | 7.7 | 6.3 | 6.8 | 11.6 |
| $z$ (RE) | 0.09960(10) | 0.09656(7) | 0.09822(12) | 0.0981(2) |
| $z$ (Bi) | 0.62381(9) | 0.62519(7) | 0.62468(13) | 0.6251(2) |
| $z$ (S1) | 0.3781(6) | 0.3789(4) | 0.3778(7) | 0.3751(11) |
| $z$ (S2) | 0.8119(4) | 0.8118(3) | 0.8143(6) | 0.8160(9) |
| $U_{iso}$ (RE) (Å$^2$) | 0.0080(5) | 0.0097(4) | 0.0084(7) | 0.0109(10) |
| $U_{11}$ (Bi) (Å$^2$) | 0.0096(4) | 0.0093(3) | 0.0102(6) | 0.0148(9) |
| $U_{33}$ (Bi) (Å$^2$) | 0.0265(11) | 0.0211(7) | 0.0254(14) | 0.022(2) |
| $U_{11}$ (S1) (Å$^2$) | 0.032(3) | 0.021(2) | 0.017(3) | 0.008(4) |
| $U_{33}$ (S1) (Å$^2$) | 0.037(6) | 0.027(4) | 0.036(7) | 0.060(10) |
| $U_{iso}$ (S2) (Å$^2$) | 0.007(2) | 0.0071(12) | 0.005(2) | 0.011(4) |